\documentclass[11pt]{article}
\usepackage{jheppub} 

\usepackage{feynmp-auto}
\usepackage{amssymb}
\usepackage{amsmath}
\usepackage{amstext}
\usepackage{graphicx,epsfig}
\usepackage{epsfig}
\usepackage{verbatim} 
\usepackage{fancyhdr}
\usepackage{fancybox}
\usepackage{color}
\usepackage{ulem,bbold}
\usepackage{enumitem}
\usepackage{caption}
\usepackage{subcaption}
\usepackage{bbm}
\usepackage{parskip}
\usepackage{appendix}
\usepackage{mathtools}

\usepackage{braket}
\usepackage{indentfirst}
\usepackage{comment}

\newcommand{\be}{\begin{equation}}
\newcommand{\ee}{\end{equation}}

\title{\boldmath On Entropy Bounds for Irrelevant Operators} 

\author[]{Lucas Fernández-Sarmiento, Riccardo Penco and Rachel A Rosen}

\affiliation[]{Department of Physics, Carnegie Mellon University, Pittsburgh, PA 15213, USA}

\emailAdd{lucasf@andrew.cmu.edu, rpenco@andrew.cmu.edu, rarosen@cmu.edu}

\abstract{
Consistency constraints for low-energy theories, especially those lacking Lorentz invariance, have recently garnered attention. Building on results from black hole thermodynamics, we investigate the conjecture that leading irrelevant deformations of a conformal field theory in the infrared must increase the system's entropy. We show that this entropy-positivity conjecture is equivalent to a decrease in the thermal grand potential at a fixed temperature. We then evaluate this proposal against various known positivity bounds and other physical constraints on effective theories: for \(U(1)\) Goldstone bosons with a quartic self-interaction at (non-)zero chemical potential, for the Euler-Heisenberg model, for the $O(N)$ nonlinear sigma model in $(2+1)D$, and for \(T\bar{T}\) deformations of the 2D Ising CFT. We find broad agreement with the entropy-positivity conjecture and we discuss test cases where the conjecture cannot be applied.
}

\begin{document} 
\maketitle
\flushbottom

\section{Introduction}
\label{sec:intro}
The standard assumption of Effective Field Theory (EFT) is that low energy theories can be constructed in a model independent way based solely on the input of the particle content and the symmetry of the low energy theory.  From this perspective, one can remain entirely agnostic about the microscopic physics. Wilson coefficients of a given theory are then taken to be fixed by experiment or observation rather than through theoretical arguments.

However, there has been great recent interest in using the principles of theoretical consistency such as causality or unitarity to constrain Wilson coefficients of low energy effective theories (see, e.g., \cite{deRham:2022hpx} and references therein).  Consistency constraints for theories without Lorentz invariance are notoriously hard to obtain.  For instance, in  \cite{Adams:2006sv} positivity bounds are derived for Lorentz invariant theories, but in general, one cannot rely on many of the analytic properties that Lorentz invariance confers when dealing with systems where this symmetry is non-linearly realized. Indeed, recent work has shown non-analyticities in the dispersion relation as a byproduct of Lorentz symmetry breaking, leading to branch cuts in unphysical regions of kinematic space \cite{Hui:2023pxc,Creminelli:2023kze}.  Nevertheless, progress has been made along a variety of axes (see, e.g., \cite{Creminelli:2022onn,Creminelli:2024lhd}).

In \cite{Cheung:2018cwt} an entropic argument was invoked to provide bounds on the coefficients of higher-dimensional operators that contribute to the entropy of a black hole. There it was shown that, for a thermodynamically stable black hole and a theory in which the dominant contributions to higher dimension operators are the tree level contributions generated by massive fields that have been integrated out, the following inequality holds at fixed temperature $T=1/\beta$:
\be
\label{DeltaF}
F(\beta) < F_0(\beta) \, ,
\ee
where $F(\beta)$ is the free energy of a theory that contains microscopic degrees of freedom that have been frozen out in the computation of $F_0(\beta)$.  For instance, $F_0$ can be taken to be the free energy of a  Reissner-Nordström black hole with a pure Einstein-Maxwell Lagrangian, while $F$ includes higher order operators.  More generally $F_0$ can refer to a theory in which the cutoff has been taken to infinity.  We emphasize that the proof of \eqref{DeltaF} relies on the fact that the dominant contribution to black hole thermodynamics comes from the saddle point, and on the assumption of thermodynamical stability, i.e. that the saddle point is a minimum of the Euclidean action.

The authors of \cite{Cheung:2018cwt} then show that the inequality \eqref{DeltaF} is equivalent to an inequality of the entropies at fixed black hole mass $M$:
\begin{equation}
    \label{constraint_free_energy_entropy}
    F(\beta)<F_0(\beta)  \implies S(M)>S_0(M) \, .
\end{equation}
These entropy inequalities give new positivity bounds on the coefficients of higher-dimension operators in gravitational theories.

In this paper, we aim to exploit the fact that the entropy constraint $\Delta S > 0$ can be applied to theories without Lorentz invariance, unlike many traditional positivity bounds. While we cannot prove $F(\beta) < F_0(\beta)$ away from the saddle point approximation, we conjecture that it must be true for irrelevant deformations of an infrared conformal field theory even when the leading contribution to the free energy comes from loops. This conjecture is a weaker statement than the claim that the thermal free energy should be monotonic along the RG flow~\cite{Appelquist:1999hr,Appelquist:1999vs,Periwal:1994im}, for which there are known counterexamples in the literature~\cite{Sachdev:1993pr,Klebanov:2011gs}. We will check the validity of our conjecture in a variety of known systems. Our assumption has physical justification. Namely, irrelevant corrections to higher dimension operators encode the presence of microscopic degrees of freedom that have been integrated out, and which presumably would give rise to a larger number of microscopic configurations corresponding to the same macroscopic state~\cite{Kats:2006xp}. Thus, we would expect a theory with these operators to have a greater entropy than a theory in which the cutoff (i.e., the mass of the heavy fields) has been taken to infinity and the heavy degrees of freedom are decoupled.

Note that the contribution to the entropy coming from the dispersion relation of the heavy modes of mass $M$  will be suppressed as $e^{-\beta M}$ compared to the contributions from irrelevant corrections to the low-energy Lagrangian, which will be polynomial in the low temperature regime $\beta M \gg 1$. 

In what follows, we show in full generality that, whenever an inequality exists for the grand potential energy $\Delta \Omega(\beta,\mu) < 0$, it implies an equivalent inequality for the entropy as a function of the extensive variables, $\Delta S (E, Q)>0$, without any saddle point approximation or assumption of Lorentz invariance.  We will then apply the bound $\Delta S >0$ to multiple systems for which the leading thermodynamic behavior doesn't come from the saddle point.  We will find that the bounds from the entropy inequality pass some nontrivial consistency checks and are compatible with known positivity arguments in all the examples we consider. 

The outline of this paper is as follows.  In section \ref{sec:EntropyandFE} we will construct a general equivalence between a negative shift in grand potential energy $\Omega(\beta, \mu)$ and a positive shift in entropy $S(E, Q)$.  In section \ref{sec:free} we will briefly discuss entropy considerations for a non-interacting theory.  In section \ref{sec:FTFT} we will consider several applications of our entropy bounds, to theories both with and without Lorentz invariance.  In section \ref{sec:cautionary}, we explore cases that fall outside the scope of our bounds, where these results may not apply. We end with some discussion in section \ref{sec:discussion}.

{\bf Conventions and Notation:}  We work in the mostly plus metric signature $(-,+,+,+)$, we adopt natural units such that $c=\hbar = k_B = 1$, and we restrict ourselves to 4 spacetime dimensions unless otherwise indicated.

\section{Entropy and Grand Potential Energy}
\label{sec:EntropyandFE}

In \cite{Cheung:2018cwt}, it was shown that a negative change in the free energy \(F\) at fixed temperature \(\Delta F(\beta) < 0\) is equivalent to a positive change in the entropy \(S\) at fixed energy \(\Delta S(E) > 0\). Here we generalize this result by considering systems at finite chemical potential \(\mu\). Thus, instead of the Helmholtz free energy \(F\), we will be using the grand potential \(\Omega \equiv -T \ln{\cal Z} = F-\mu \, Q\).

A known subtlety with the grand potential (and free energy) in quantum field theory is its sensitivity to UV divergences which are independent of temperature, such as the zero-point energy. To construct an argument insensitive to such effects, we define the ``thermal" part of the grand potential as\
\begin{equation}
\label{thermalOm}
    \Omega^*(T,\mu) \equiv \Omega(T,\mu) - \Omega(T=0,\mu).
\end{equation}
This quantity is UV finite and vanishes as \(T \to 0\). Correspondingly, we define the thermal energy \(E^*(T,\mu) = E(T,\mu) - E(T=0,\mu)\) and thermal charge \(Q^*(T,\mu) = Q(T,\mu) - Q(T=0,\mu)\).\footnote{We note that, because of the Ward identity associated with the conserved current, one can show that the conserved current is not subject to infinite renormalization, even in cases of spontaneous symmetry breaking (see, e.g., \cite{Cheng:1984vwu}).  However, here we keep $Q^*$ defined for notational convenience.} The entropy \(S(T,\mu)\) is inherently a thermal quantity, satisfying \(S(T=0,\mu)=0\) by the third law of thermodynamics (assuming a non-degenerate ground state). These quantities are related by \(\beta \Omega^* = \beta E^* - S - \beta \mu Q^*\), which is analogous to the standard relation for \(\Omega\).

In what follows, we will show that a negative change in these thermal grand potentials at fixed temperature and chemical potential, \(\Delta \Omega^*(\beta,\mu) < 0\), is equivalent to a positive change in entropy at fixed thermal energy \(E^*\) and thermal charge \(Q^*\): \(\Delta S(E^*,Q^*)>0\). When \(\mu = 0\), then \(\Omega^* = F - F(T=0)\), and our result reduces to a temperature-subtracted version of that in \cite{Cheung:2018cwt}.

Our starting point is two distinct theories: a ``reference theory" whose thermal grand potential is given by \(\Omega_0^*(\beta_0,\mu_0)\) and a ``target theory" whose thermal grand potential is given by \(\Omega^*(\beta,\mu)\). In particular, we consider a target theory and a reference theory such that when their temperatures and chemical potentials are matched, i.e., \(\beta_0 = \beta\) and \(\mu_0 = \mu\), the following inequality holds:
\begin{equation}
 \label{grandinequality_mod}
    \beta\,\Omega^*(\beta, \mu)< \beta\, \Omega_0^*(\beta,\mu) \, .
\end{equation}
Let us now suppose we take our two theories and fix the thermal energy of both to be \(E^*\) and the thermal charge of both to be \(Q^*\). In general, this will correspond to different values of temperature and chemical potential for each theory. Let \((\beta, \mu)\) be the parameters for the target theory that yield \(E^*\) and \(Q^*\), and let \((\beta_0, \mu_0)\) be the parameters for the reference theory that yield the same \(E^*\) and \(Q^*\). If the theories are perturbatively close such that these parameters are close, we can write
\begin{equation}
    \beta=\beta_0+\Delta \beta\, , ~~~~
    \mu=\mu_0+\Delta \mu \, .
\end{equation}
Let us now consider the right hand side of equation \eqref{grandinequality_mod}, where \((\beta, \mu)\) are the parameters of the target theory. We expand this term around the parameters \((\beta_0, \mu_0)\) of the reference theory, keeping in mind that \(E_0^*(\beta_0,\mu_0) = E^*\) and \(Q_0^*(\beta_0,\mu_0) = Q^*\):
\begin{align}
    \beta\, \Omega_0^*(\beta,\mu)&\simeq\beta_0\Omega_0^*(\beta_0,\mu_0)
    +\Delta \beta\,\left. \frac{\partial}{\partial \beta'}\left[\beta' \,\Omega_0^*(\beta',\mu')\right] \right|_{\beta_0,\mu_0}
     +\Delta \mu\,\left. \frac{\partial}{\partial \mu'}\left[\beta' \,\Omega_0^*(\beta',\mu')\right] \right|_{\beta_0,\mu_0} \nonumber \\
    &=\beta_0 \Omega_0^*(\beta_0,\mu_0)+\Delta \beta\, (E^*-\mu_0 Q^*)-\Delta \mu\, \beta_0 Q^* \, . \label{expansion_mod} 
\end{align}
To get the last lines, we have used the thermodynamic relations for the starred quantities:
\begin{align}
\left. \frac{\partial}{\partial \beta'}\left[\beta' \,\Omega_0^*(\beta',\mu')\right] \right|_{\beta_0,\mu_0} &= \Omega_0^*(\beta_0,\mu_0) + \beta_0 \left. \frac{\partial}{\partial \beta'}\left[\Omega_0^*(\beta',\mu')\right] \right|_{\beta_0,\mu_0} \nonumber  \\ &= \Omega_0^* (\beta_0,\mu_0)+ \frac{S_0(\beta_0,\mu_0)}{\beta_0} \nonumber\\
&= E^*-\mu_0 Q^* \ , \\
\nonumber\\
 \left. \frac{\partial}{\partial \mu'}\left[\beta' \,\Omega_0^*(\beta',\mu')\right] \right|_{\beta_0,\mu_0} &= \beta_0 \left. \frac{\partial}{\partial \mu'}\left[\Omega_0^*(\beta',\mu')\right] \right|_{\beta_0,\mu_0} \nonumber\\
&= - \beta_0 Q^* \ .
\end{align}

We now plug the expansion \eqref{expansion_mod} on the right-hand side of the inequality \eqref{grandinequality_mod}. Using \(\beta \Omega^* = \beta E^* - S - \beta \mu Q^*\) on the left-hand side, we can write the inequality (evaluated at fixed \(E^*, Q^*\)) as
\begin{equation}
    \beta E^* - S(E^*,Q^*) - \beta \mu Q^* < \beta_0 E^* - S_0(E^*,Q^*) - \beta_0 \mu_0 Q^* + \Delta \beta\, (E^*-\mu_0 Q^*) - \Delta \mu\, \beta_0 Q^* \, .
\end{equation}
Combining terms and canceling on both sides (where we are still working to first order only in small perturbations \(\Delta \beta\) and \(\Delta \mu\)), we find \(-S(E^*,Q^*) < -S_0(E^*,Q^*)\). In other words
\begin{equation}
 \Delta S(E^*,Q^*) \equiv S(E^*,Q^*) - S_0(E^*,Q^*) > 0    \, .
\end{equation}
It is important to note that this inequality for entropies holds at equal thermal charge \(Q^*\) and thermal energy \(E^*\), while the inequality \eqref{grandinequality_mod} relating thermal grand potentials is at equal temperature and chemical potential. This equivalence relies only on general thermodynamic relations and is therefore insensitive to temperature-independent UV divergences, provided \(\Omega^*(T,\mu)\) is well-defined. 

The central physical assumption lies in the initial inequality, $\Delta\Omega^*(\beta,\mu) < 0$. In the rest of this paper, we will investigate to what extent this inequality is satisfied when the reference theory is a conformal field theory at an IR fixed point and the target theory includes an irrelevant deformation of the IR theory.  As noted in the introduction, we cannot {\it prove} that the inequality \ref{grandinequality_mod} holds for any arbitrary deformation.  However, we posit that the leading irrelevant operators will consistently yield a negative correction to the thermal grand potential, and by extension, a positive correction to the entropy and we present a range of examples that support this conjecture.  The gravitational proof of $\Delta F < 0$ derived in~\cite{Cheung:2018cwt} is a statement about the full free energy and not the thermal free energy.  From this perspective, our conjecture is a generalization of this physical principle from the gravitational saddle-point context to thermal quantum field theories more broadly.

\section{Thermal Field Theory Review}
\label{sec:free}

Before proceeding, we first review some basic results concerning the grand potential energy of a  quantum field theory at finite temperature.  Our summary follows closely the presentation of \cite{Kapusta:2006pm}.  In particular, the grand potential energy is given by 
\begin{equation}
    \Omega =-T \ln {\cal Z} \, ,
\end{equation}
where ${\cal Z}$ is the partition function defined by
\begin{equation}
   {\cal Z} = N \int \mathcal D\phi\, e^{-S_E[\phi]}\,,
  \qquad
  S_E = \int_{0}^{\beta} d\tau \int d^{3}x \mathcal L_E(\phi)\,.
  \label{eq:Z-def}
\end{equation}
and $N$ is an overall (irrelevant) normalization constant.

It is useful to split the action $\cal S$ into a piece ${\cal S}_{\rm free}$ that is at most quadratic in the fields and a piece $S_{\rm int}$ that is higher order: ${\cal S}={\cal S}_{\rm free} +{\cal S}_{\rm int}$.  The contribution to the grand potential energy coming from ${\cal S}_{\rm free}$ can be computed exactly.  For a single field with dispersion relation $\omega(k)$, it is given by\footnote{This can be trivially extended to multiple fields by summing the second and third terms over the dispersion relations of each field.}
\be
\label{Omegabar}
\Omega_{\rm free} = V \langle {\cal L}_E \rangle+V \int \frac{d^3 k}{(2\pi)^3}\,\omega(k) +T\,V \int \frac{d^3 k}{(2\pi)^3} \ln\left[1 - e^{-\beta \omega(k)}\right] \, .
\ee
Here $V$ is the volume, and $\langle \mathcal{L}_E \rangle$ is the volume average of the Euclidean Lagrangian density evaluated on the background value of the field.  The first term is the saddle point contribution. The second term is the zero-point energy of the field. Finally, the third term is the only one that is temperature-dependent and arises from thermal fluctuations of the field. Note that this term generically also depends on the chemical potential through the dispersion relation~$\omega (k)$.

In what follows, we will be interested only in the thermal grand potential energy \eqref{thermalOm}:
\be \label{eq: thermal free energy free (3+1)D}
\Omega^*_{\rm free} = T\,V \int \frac{d^3 k}{(2\pi)^3} \ln\left[1 - e^{-\beta \omega(k)}\right] \, .
\ee
Let us compute this explicitly for a simple Lorentz-violating dispersion relation that will be useful later.  In particular, we consider a general dispersion relation of the form
\be 
\label{disp}
\omega(k) = c_s k+\frac{\alpha}{\Lambda^2} k^3 + \dots \, ,
\ee
where we have dropped higher order corrections in $k$ because they would be negligible at sufficiently small $k$. Integrating out massive modes around a background that breaks Lorentz spontaneously, e.g. because of a non-zero chemical potential $\mu$, will generically give rise to $c_s \neq 1$ and $\alpha \neq 0$. In this case, $c_s$ and $\alpha$ would depend on $\mu$. Expanding in $k/\Lambda$, the thermal grand potential energy arising from the dispersion relation \ref{disp} is given by
\be
\label{barOmstar}
\Omega_{\rm free}^* \simeq -\frac{\pi^2}{90\, c_s^3\, \beta^4}\, V
+\frac{\alpha}{\Lambda^2} \frac{4 \pi^4}{63\, c_s^6\, \beta^6}\, V \, .
\ee
To take into account the terms in ${\cal S}_{\rm int}$ which are higher than quadratic order in the fields, the full machinery of perturbative thermal quantum field theory is invoked~\cite{Kapusta:2006pm}.  In the following sections we will employ this machinery to calculate the contribution of such terms to the thermal grand potential energy for a variety of models.

\section{Constraining Finite Temperature Field Theories}
\label{sec:FTFT}
We now consider several examples of finite temperature field theories, both with and without Lorentz invariance.  We verify that our entropy bound is consistent with the positivity bounds already known in the literature for these theories or, in cases where the Wilson coefficients are already known, provide non-trivial consistency checks.

\subsection{$U(1)$ Goldstone with quartic self-interaction}
\label{subsec:dphi4}
We start by applying our entropy bound to the case of the well-studied $(\partial \phi)^4$ theory, i.e., the quartic goldstone. The Lagrangian is that of a free massless scalar with a higher dimensional term suppressed by the scale $M$.
\begin{equation}
S = \int d^4 x \left[ -\frac{1}{2}\partial^\mu \phi \partial_\mu \phi+\frac{\lambda}{M^4}\left(\partial_\mu \phi \partial^\mu \phi\right)^2 \right]. 
\label{dphi4}
\end{equation}
We can think of $M$ as the mass of some heavy field that we have integrated out to obtain this effective action. Famously, it has been shown by using positivity bounds as well as by invoking the absence of superluminality around non-trivial backgrounds in the low-energy theory that the value of $\lambda$ must be greater than zero \cite{Adams:2006sv}. Here, we show how the entropy bound yields the same result in two distinct cases: first in which the background is Lorentz invariant and then in the case that Lorentz invariance is broken by a chemical potential.

\subsubsection{Lorentz-Invariant Background}
\label{subsubsec:dphi4LI}
We first consider the $(\partial \phi)^4$ theory around a Lorentz invariant background. We will thus treat the free theory as our reference theory, and use standard methods of quantum field theory at finite temperature to calculate the entropy corrections due to the $(\partial \phi)^4$ interaction.
\begin{figure}[t!]
\begin{subfigure}{.5\textwidth}
  \centering
  \epsfig{file=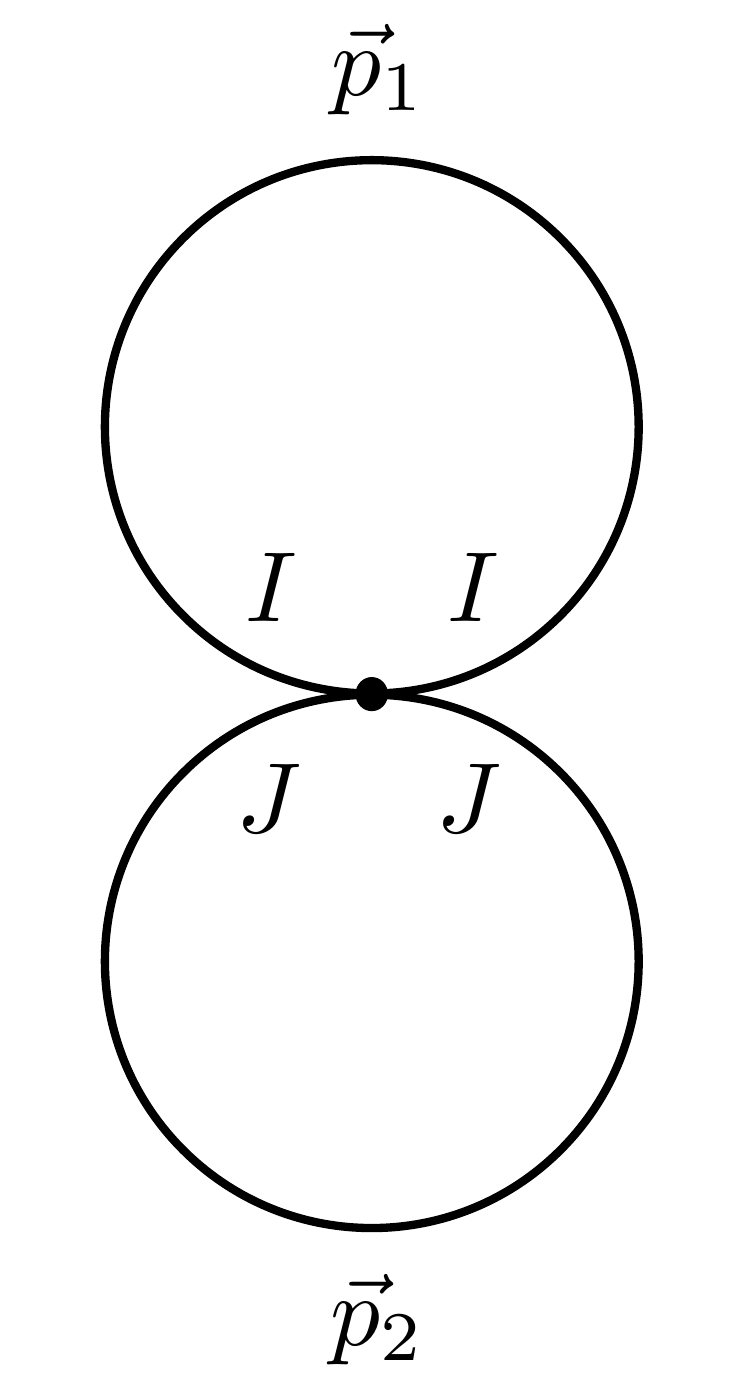,width=0.55\linewidth}
  \caption{Diagram A}
  \label{fig:diaga}
\end{subfigure}
\begin{subfigure}{.5\textwidth}
  \centering
\epsfig{file=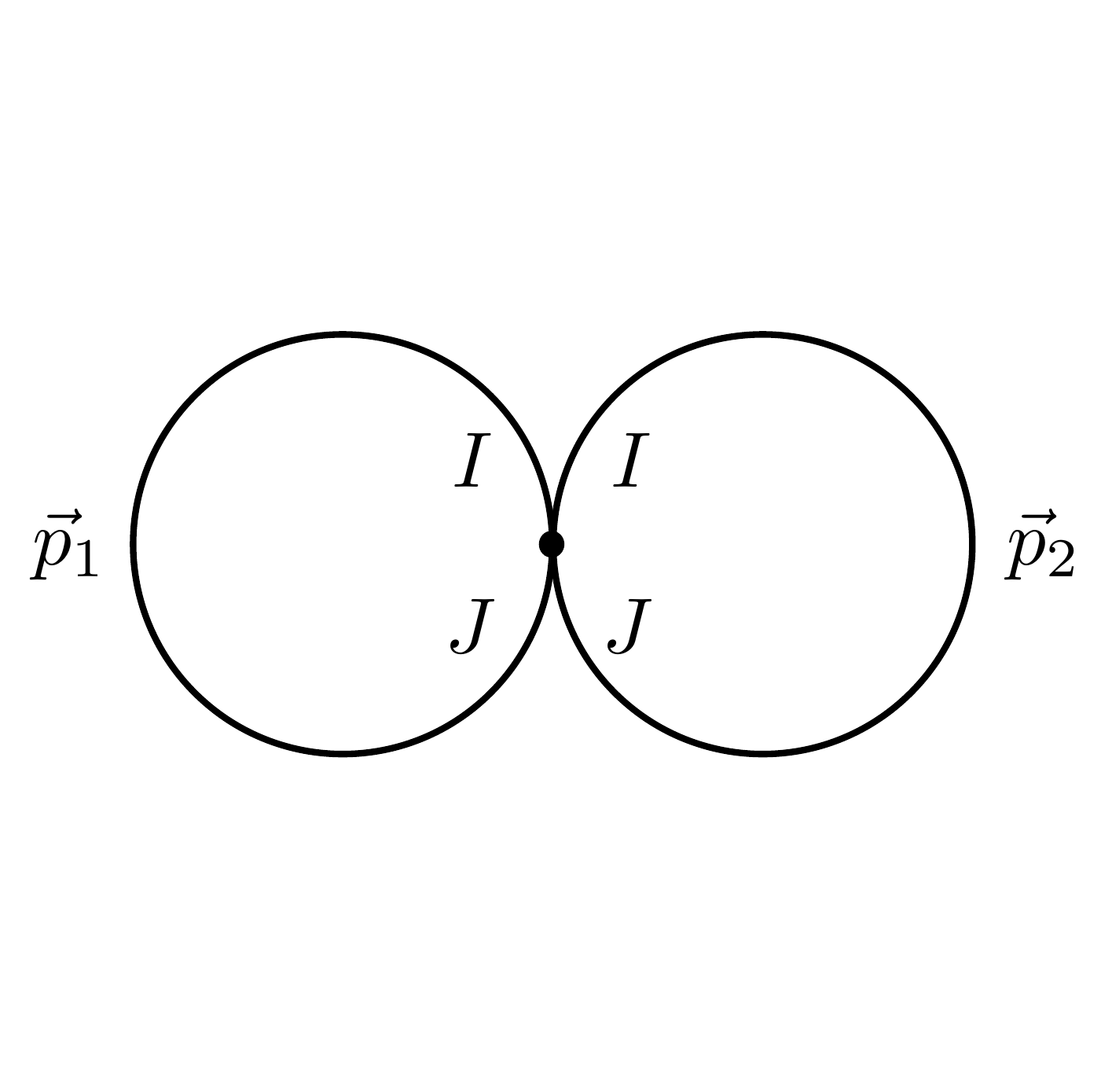,width=1\linewidth} 
  \caption{Diagram B}
  \label{fig:diagb}
\end{subfigure}
\caption{Diagrams that contribute to the leading order correction $\Delta\Omega$ to the grand potential energy  due to a $(\partial \phi)^4$ interaction.}
\label{fig:diagrams}
\end{figure}

Based on Eq. \eqref{barOmstar} with $c_s=1$ and $\alpha = 0$, the thermal grand potential energy for the reference theory is
\be
\Omega_0^* =  -\frac{\pi^2}{90 \beta^4} V \, .
\ee
The quartic self-interactions correct the grand potential by an amount $\Delta \Omega$ which, at lowest order in perturbation theory, is described by the Feynman diagrams shown in Figure \ref{fig:diagrams}. Here, $I,J$ are Euclidean 4-vector indices such that $p^I = (\omega_n,\vec{p}\,)$ and $\omega_n = 2\pi n T$.  The second diagram has twice as many contractions and thus comes with an extra factor of two relative to the first one. The leading order correction to the grand potential energy thus takes the form
\be
\label{DeltaOmega}
\Delta \Omega = - 24 \frac{\lambda}{M^4} V \times \left( X_A+ X_B\right) \, ,
\ee
where $X_A$ is the contribution from Diagram A,
\be
\label{XA}
   X_A=  \left(T \sum_n \int \frac{d^3 p}{(2 \pi)^3} \frac{p^I p^I}{\omega_n^2+\vec{p}^{\,2}}\right)^2
    = \left(T \sum_n \int \frac{d^3 p}{(2 \pi)^3} \frac{\omega_n^2+\vec{p}^{\,2}}{\omega_n^2+\vec{p}^{\,2}}\right)^2
    = \left(T \sum_n \int \frac{d^3 p}{(2 \pi)^3} \right)^2 \, ,
\ee
and $X_B$ is the contribution from the Diagram B,
\be
\label{XB}
   X_B = 2 \left(T \sum_{n_1} \int \frac{d^3 p_1}{(2 \pi)^3} \frac{p_1^I
   \,p_1^J}{\omega_{n_1}^2+\vec{p}_1^{\,2}}\right) \left(T\sum_{n_2}\frac{d^3 p_2}{(2 \pi)^3}
  \frac{p_2^I\, p_2^J}{\omega_{n_2}^2+\vec{p}_2^{\,2}}\right) \, ,
\ee
with $p_1^I\, p_2^I = \omega_{n_1}\omega_{n_2} + \vec{p_1} \cdot \vec{p}_2 $.  Both $X_A$ and $X_B$ are UV divergent.  In what follows, we will show that the UV divergences of both terms at finite temperature are identical to the UV divergences at zero temperature.  Since we are only interested in finite temperature quantities, we will subtract off the the zero temperature contributions.  We find that the finite temperature contribution coming from \eqref{XA} is zero while the contribution coming from \eqref{XB} is finite.

Before calculating these contributions explicitly, however, we note that we can make the following general argument.  From symmetry, we know that terms linear in the three momenta must vanish:
\be
\int \frac{d^3 p}{(2 \pi)^3} \frac{\vec{p}}{\omega_{n}^2+\vec{p}^{\,2}} = 0 \, .
\ee
Therefore, terms proportional to, say, $\omega_{n_1} \,\vec{p}_1$ in the numerator of the integrand will be zero.  The only terms that contribute to the grand potential energy are thus the terms that are perfect squares 
\be
\label{DeltaOmegasimp}
\Delta \Omega = -24 \frac{\lambda}{M^4} V \times
\left[ \left(T \sum_n \int \frac{d^3 p}{(2 \pi)^3}\right)^2
   +2 T^2 \sum_{n_1}\sum_{n_2} \int \frac{d^3 p_1}{(2 \pi)^3} \frac{d^3 p_2}{(2 \pi)^3} \frac{\omega_{n_1}^2\,\omega_{n_2}^2+(\vec{p}_1 \cdot \vec{p}_2)^2}{(\omega_{n_1}^2+\vec{p}_1^{\,2})(\omega_{n_2}^2+\vec{p}_2^{\,2})} \right] .
\ee
Thus the shift in the grand potential energy $\Delta \Omega$ due to the $(\partial \phi)^4$ term is manifestly negative for positive $\lambda$.  It is then tempting to conclude that, as long as the coupling of the $(\partial \phi)^4$ term is positive, the change in entropy of the theory due to microscopic degrees of freedom that have been integrated out is also positive, as we would expect. This argument, however, is a bit too quick because $\Delta \Omega$ is UV divergent. 

To be more precise, let us compute $\Delta \Omega$ explicitly and extract its finite, temperature-dependent part.  To do so, we make use of the identity given in \cite{Kapusta:2006pm}, 
\begin{equation}
\label{Kapusta}
\begin{aligned}
 T \sum_n f\left(p_0=i \omega_n\right)=& \frac{1}{2 \pi i} \int_{-i \infty}^{i \infty} d p_0 \frac{1}{2}\left[f\left(p_0\right)+f\left(-p_0\right)\right]  \\
& +\frac{1}{2 \pi i} \int_{-i \infty+\epsilon}^{i \infty+\epsilon} d p_0\left[f\left(p_0\right)+f\left(-p_0\right)\right] \frac{1}{\mathrm{e}^{\beta p_0}-1} .
\end{aligned}
\end{equation}
The first term on the right hand side of the above expression is manifestly independent of temperature.  It also contains all of the UV divergences of $\Delta \Omega$.  In other words, the UV divergences of the theory at finite temperature are the same as that at zero temperature.  To extract the physical contribution to $\Delta \Omega$ we focus on the finite temperature contributions.

To calculate $X_A$, we trivially identify $f(p_0) =1$.  Substituting this into \eqref{Kapusta}, we note that the finite temperature piece contains no poles when we close the integration contour to the right.  Thus, the contribution of this piece to the finite temperature grand potential energy is zero. For $X_B$, there are two types of terms we need to evaluate:
\be
\label{interest}
T \sum_n \int \frac{d^3 p}{(2 \pi)^3} \frac{\omega_n^2}{\omega_n^2+\vec{p}^{\,2}} \, ,
~~~~{\rm and} ~~~~
T \sum_n \int \frac{d^3 p}{(2 \pi)^3} \frac{p^i p^j}{\omega_n^2+\vec{p}^{\,2}} \, .
\ee
For the first term of interest in \eqref{interest}, we identify 
\be
\label{fp0}
f(p_0)=\frac{-p_0^2}{-p_0^2+\vec{p}^{\,2}} \, .
\ee
Then, using the identity \eqref{Kapusta}, we can express the temperature dependent piece as
\be
T \sum_n \int \frac{d^3 p}{(2 \pi)^3} \frac{-p_0^2}{-p_0^2+\vec{p}^{\,2}}  = {\cal O}(T^0) +\frac{1}{2\pi i}\int \frac{d^3 p}{(2 \pi)^3} \int_{-i \infty+\epsilon}^{i \infty+\epsilon} d p_0 \frac{2p_0^2}{p_0^2-\vec{p}^{\,2}} \frac{1}{\mathrm{e}^{\beta p_0}-1}
\ee
To carry out the integral over $p_0$, we can close the contour on the right side of the complex plane, encircling the $p_0=\sqrt{\vec{p}^{\,2}} \equiv p$ pole on the real axis. We pick up a minus sign as we travel clockwise along the contour so that, using Cauchy's theorem, we find 
\begin{equation}
+\frac{1}{2\pi i}\int_{-i \infty+\epsilon}^{i \infty+\epsilon} d p_0 \frac{2p_0^2}{p_0^2-\vec{p}^{\,2}} \frac{1}{\mathrm{e}^{\beta p_0}-1}=-\frac{p}{\mathrm{e}^{\beta p}-1}.
\end{equation} 
Thus, for the finite temperature piece of the first expression in \eqref{interest} we find
\begin{equation}
  \begin{aligned}
    T \sum_n \int \frac{d^3 p}{(2 \pi)^3} \frac{\omega_n^2}{\omega_n^2+\vec{p}^{\,2}} 
   &= {\cal O}(T^0) -\int \frac{d^3 p}{(2 \pi)^3}\frac{p}{\mathrm{e}^{\beta p}-1} \\
   &= 
   {\cal O}(T^0)-\frac{\pi^2}{30\beta^4} \, .
   \end{aligned}
\end{equation}

We can compute the contribution from the second term in \eqref{interest}, i.e., the $p^i p^j$ terms following a similar procedure.  Based on symmetry arguments, we know that
\begin{equation}
T \sum_n\int \frac{d^3 p}{(2 \pi)^3} \frac{p^i p^j}{\omega_n^2+\vec{p}^{\,2}}=A\, \delta^{ij}
\end{equation}
due to the lack of any other covariant tensor structures.  Contracting both sides of the above expression with $\delta_{ij}$, we identify
\be
f(p_0)=\frac{\vec{p}^{\,2}}{-p_0^2+\vec{p}^{\,2}} \, .
\ee
On the $p_0=\sqrt{\vec{p}^{\,2}}$ pole, this function has the opposite sign as \eqref{fp0}. Following the same steps as above for this choice of $f(p_0)$ yields 
\be
A=+\frac{\pi^2}{90\beta^4} \, .
\ee
where again we are neglecting the zero temperature contribution.

Putting everything together in \eqref{DeltaOmegasimp} now gives the thermal contribution
\be
\Delta \Omega^* = -48 \frac{\lambda}{M^4} V \times \left[ \left(-\frac{\pi^2}{30\beta^4}\right)^2+\delta^{ij} \delta_{ij}\left(\frac{\pi^2}{90\beta^4}\right)^2 \right] 
= -\frac{\lambda}{M^4} \frac{16 \pi^4}{225 \beta^8} V \, ,
\ee
where $\delta^{ij} \delta_{ij} = 3$.  We see explicitly that, for positive $\lambda$, the change in the grand potential energy due to the $(\partial \phi)^4$ term is negative and thus, based on the arguments in Sec. \ref{sec:EntropyandFE}, the change in the entropy at constant energy is positive, as expected.

\subsubsection{Finite Chemical Potential}
It is also instructive to consider the $(\partial \phi)^4$ theory in the presence of a small chemical potential, $\mu \ll M$, for the $U(1)$ shift symmetry. This is equivalent to expanding $\phi$ around the Lorentz-violating background~\cite{Nicolis:2011pv}.
\be
\label{background}
\phi=  M \mu t +\pi\, .
\ee
Then, the effective action \eqref{dphi4} gives rise to the following quadratic Lagrangian for $\pi$
\be
{\cal L}^{(2)} =\frac{1}{2} \left(1+12 \lambda \frac{\mu^2}{M^2}\right) \dot{\pi}^2 - \frac{1}{2} \left(1+4 \lambda \frac{\mu^2}{M^2}\right)(\nabla \pi)^2 \, .
\ee
After canonically normalizing $\pi$, we can read off the sound speed as
\be
\label{cs}
c_s=1-4 \lambda \frac{\mu^2}{M^2} \, .
\ee
Insisting on the absence of superluminality around this background, we would require $\lambda>0$ in keeping with the usual positivity bounds.

Following the arguments of section \ref{sec:free}, $\lambda >0 $ is also in keeping with a positive change in the entropy due to the presence of additional degrees of freedom.  In contrast to the Lorentz invariant case, however, the $(\partial \phi)^4$ interaction shows up already at quadratic order in the fields $\pi$, and therefore it now also yields a contribution proportional to~$T^4$. In the regime $\mu \gg T^2/M$, this correction dominates over the one we calculated in the previous section.  Treating again $\lambda$ perturbatively, the leading correction to the grand potential energy in this regime can be easily obtained from \eqref{barOmstar}:
\be
\Delta \Omega^* \simeq -\left(4 \lambda \frac{\mu^2}{M ^2}\right)\left( \frac{\pi^2}{30 \beta^4}V \right) \, .
\ee
Once again we see that $\lambda >0$ is consistent with a negative change to the grand potential energy and thus a positive change in the entropy.
 
 What's notable is that, while the entropy bound gives the same result, namely $\lambda>0$, in both the Lorentz invariant and Lorentz violating cases of the $(\partial \phi)^4$ theory, the mechanism is distinct. In particular the leading order corrections to the grand potential energy and thus the entropy are suppressed by $M^4$ in the Lorentz invariant case while they are suppressed by $M^2$ for the Lorentz violating case. 

\subsection{Euler-Heisenberg theory}
\label{subsec:EH}
\begin{figure}[t!]
\begin{center}
  \epsfig{file=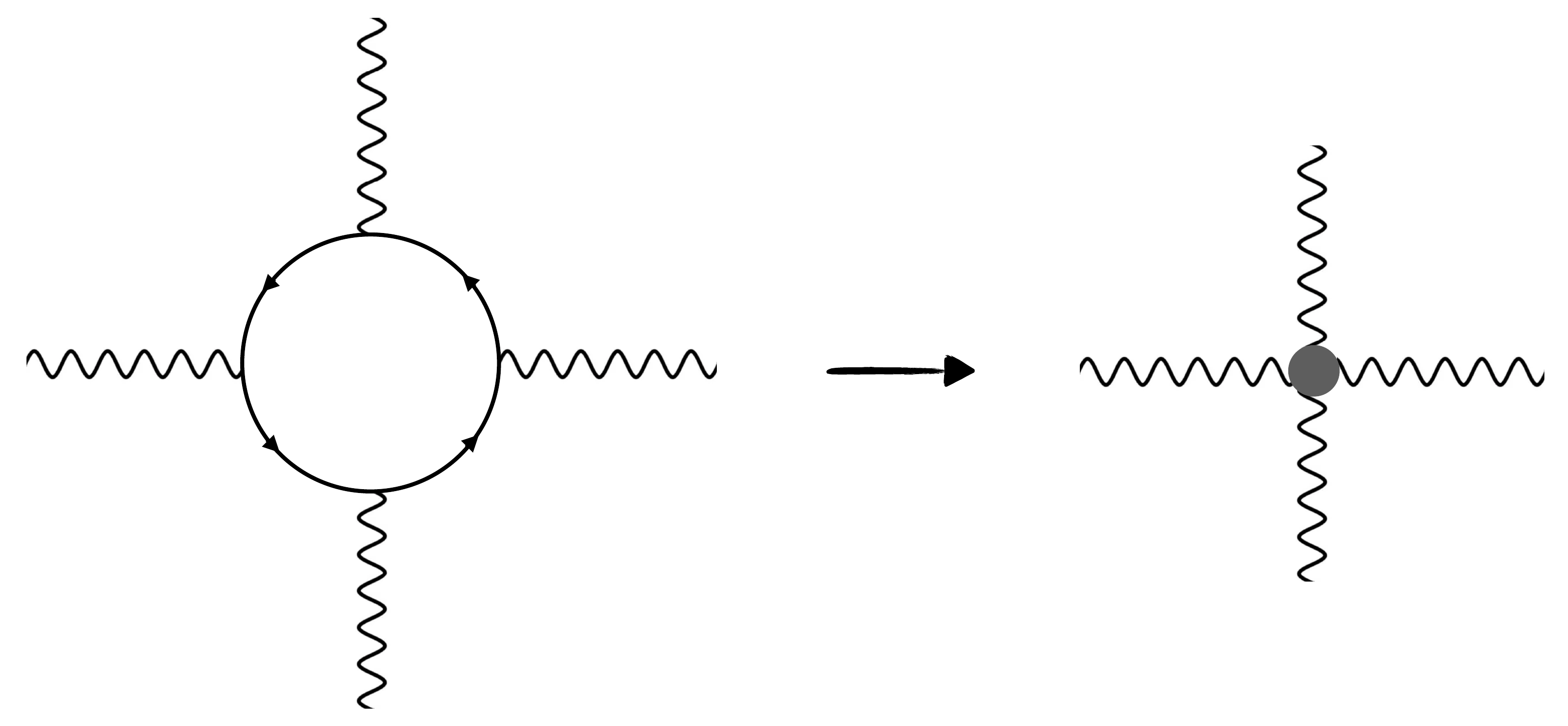,width=0.75\linewidth}
  \caption{The Euler-Heisenberg terms from integrating out massive fields}
   \label{fig:EH} 
  \end{center}
\end{figure}

We next consider the Euler-Heisenberg Lagrangian \cite{Heisenberg:1936nmg,Dunne:2004nc} that arises when integrating out heavy degrees of freedom in QED or scalar QED. This theory contains an $F^4$ term due to these massive fields in addition to the usual kinetic term in Maxwell theory.  The general action is given by\footnote{We note that a third possible quartic structure $F_\alpha^{~\beta}F_\beta^{~\gamma}F_\gamma^{~\delta}F_\delta^{~\alpha}$ is redundant with the other two.}
\begin{equation}
    S=\int d^4x \left[-\frac{1}{4}F^{\mu\nu}F_{\mu\nu}+\frac{c_1}{M^4}\, (F^{\mu\nu}F_{\mu\nu})^2+\frac{c_2}{M^4}\, (F^{\mu\nu}\tilde{F}_{\mu\nu})^2\right] \, ,
    \label{f4Lagrangian}
\end{equation}
where $\tilde{F}_{\mu\nu} = \tfrac{1}{2!} \epsilon_{\mu\nu\alpha\beta} F^{\alpha \beta}$.  We expect $c_1, c_2$ to go like $\sim \alpha_{em}^2$ where $\alpha_{em}$ is the usual fine structure constant, as these terms are generated by integrating loops as such shown in Figure \ref{fig:EH}, containing 4 massive propagators and 4 vertices.  

In particular, we can start from the UV theory
\begin{equation}
\mathcal{L}_{\mathrm{QED}}=-\frac{1}{4} F_{\mu \nu} F^{\mu \nu}+\bar{\psi}\Big(-i \gamma^\mu\left(\partial_\mu+i e A_\mu\right)-m_e\Big) \psi \, .
\end{equation}
Integrating out the electron leads to the Euler-Heisenberg Lagrangian \eqref{f4Lagrangian} with specific coefficients (see, e.g., \cite{Meissner2012})
\be
c_1=\frac{\alpha^2_{em}}{90}\,, ~~~~ c_2=\frac{7\alpha^2_{em}}{760}\, ,
\ee
and with $M \rightarrow m_e$.  Both of these coefficients are positive.  
That each coefficient be positive independently is required by the analyticity arguments of \cite{Adams:2006sv}. We will now consider analogous constraints from an entropy perspective. 

The full calculation of the grand potential energy due to the higher derivative terms is straightforward and follows very closely that of section \ref{subsubsec:dphi4LI}.  The final contribution of the higher derivative terms to the grand potential energy is given by
\begin{equation}
      \Delta \Omega^* = -\frac{c_1+c_2}{M^4}\frac{128 \pi^4 }{225 \beta^8} V \, .
\end{equation}
We can thus use our entropic considerations to argue that $c_1+c_2 >0$.

\subsection{$O(N)$ Nonlinear sigma model in $(2+1)D$}
\label{subsec:ON}

The next example of a Lorentz invariant theory we will consider is the $O(N)$ nonlinear sigma model in 2+1 dimensions. The effective action for this theory is given by \cite{Hofmann:2009ru}
\begin{equation}
S = \int d^3x \left[ - \frac{f}{2}  \, \partial_\mu \vec \Phi \cdot \partial^\mu \vec \Phi  + \mathcal{O}(\partial^4) \right]\ ,
\label{nlsmaction}
\end{equation}
where $\vec \Phi$ is an $N$-dimensional real field satisfying $\vec \Phi \cdot \vec \Phi = 1$, $f$ is the spin stiffness (with dimension of mass in 2+1 dimensions), and higher derivative terms are suppressed by a scale that is parametrically of order $f$.

This model provides a particularly stringent test of our entropy bound because, unlike the cases considered thus far, the leading correction to the thermal grand potential energy is completely determined by symmetries alone for $N>2$. In particular, as shown in \cite{Hofmann:2009ru}, the interaction contribution to the thermal grand potential energy density at order $T^5$ depends only on the spin stiffness $f$ from the leading order effective Lagrangian and is independent of any Wilson coefficients from higher derivative corrections. This means there are no adjustable parameters that could be tuned to satisfy our entropy bound—the theory either passes or fails this consistency check based purely on its symmetry structure.

Borrowing the results of \cite{Hofmann:2009ru} in the limit where all explicit symmetry breaking terms are neglected, we find that the thermal grand potential energy density for the $O(N)$ antiferromagnet is given by

\begin{figure}[t]
    \centering
    \includegraphics[width=0.75\linewidth]{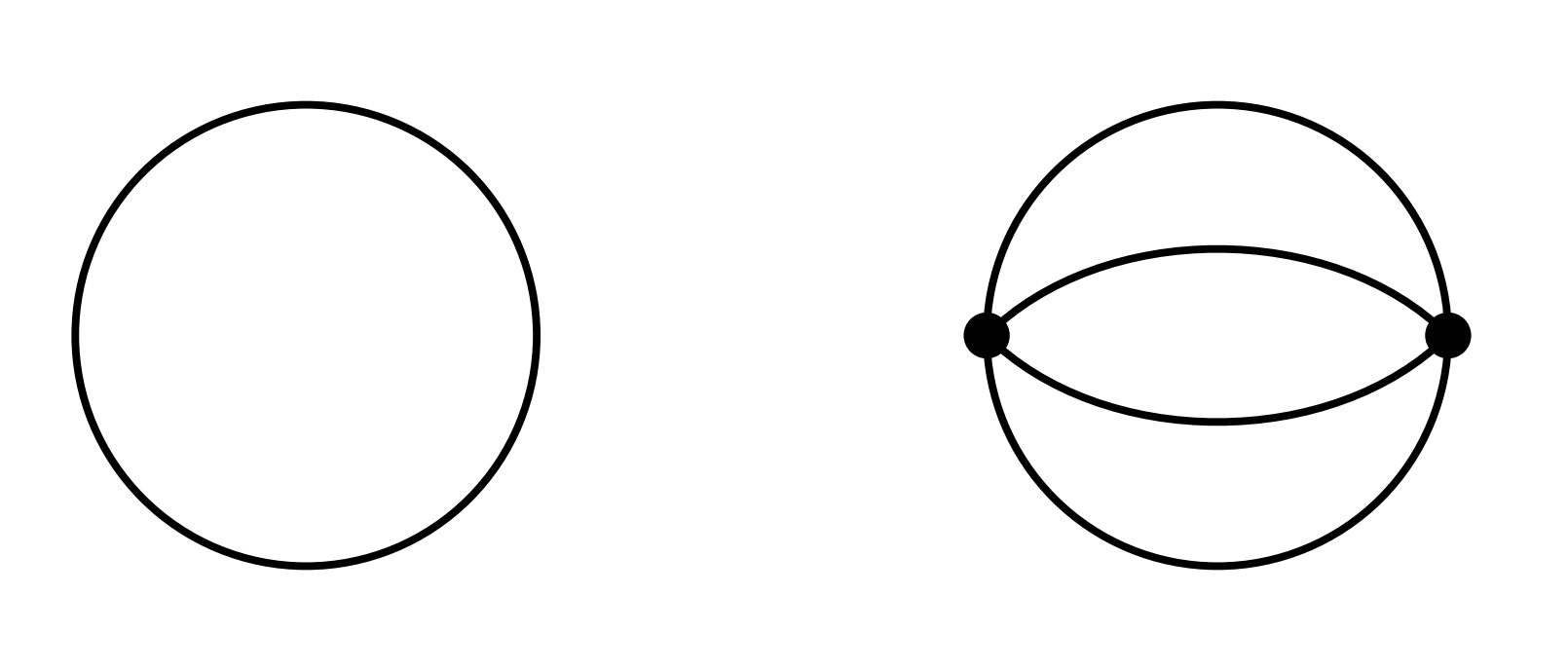}
    \caption{Diagrams contributing to the $(2+1)D$ $O(N)$ nonlinear sigma model up to $O(1/f^2)$ in the massless limit. All the vertices come from expanding the leading action in \ref{nlsmaction}.
    }
    \label{fig:enter-label}
\end{figure}

\begin{equation} \label{eq: grand potential O(N) model}
\frac{\Omega^*}{V}= -\frac{1}{2}(N-1)\frac{\zeta(3)}{\pi} T^3 + \frac{q_1}{2f^2}(N-1)(N-2) T^5 + O(T^6),
\end{equation}
where $\zeta(3) \approx 1.202$ is the Riemann zeta function and $q_1$ is a calculable numerical coefficient. The first term represents the free Bose gas contribution from $(N-1)$ Goldstone bosons, while the second term encodes the leading interaction effects. 

To extract the value of $q_1$, we use the fact that for $N=3$, the thermal grand potential energy density reduces to
\begin{equation}
\frac{\Omega^*}{V} = -\frac{\zeta(3)}{\pi} T^3 + \frac{q_1}{f^2} T^5 + O(T^6) \quad (N=3).
\end{equation}
From the numerical evaluation of the three-loop diagrams in \cite{Hofmann:2009ru}, one finds $q_1 = -0.008779$. The negative sign indicates that the magnon-magnon interaction is repulsive in this system.

Returning to general $N$, we see that the change in thermal grand potential energy due to the interaction terms is 
\begin{equation}
\frac{\Delta \Omega^*}{V} = \frac{q_1}{2f^2}(N-1)(N-2) T^5.
\end{equation}
Since $q_1 < 0$, we have $\Delta \Omega^* < 0$ for all $N > 2$, which implies $\Delta S > 0$ according to our result in Sec. \ref{sec:EntropyandFE}. This is precisely what we expect: the presence of interaction terms arising from integrating out heavy degrees of freedom increases the entropy of the system.

At this point, we should briefly discuss in which regime the above result is applicable in light of the Mermin-Wagner theorem. The absence of spontaneous symmetry in 2D systems at finite temperature implies that, in this model, the would-be Goldstone modes develop a gap $m$. In the limit where $N T / f \ll 1$, this gap is parametrically of the form
\begin{align}
	m \sim f e^{- 2\pi f / (TN)} \ ,
\end{align}
and is thus exponentially suppressed~\cite{Chubukov:1993aau,Hofmann:2009ru}. In this regime, the dominant correction to the free energy can be obtained by treating the particles as massless.

The $O(N)$ nonlinear sigma model in 2+1 dimensions is also an interesting example because it illustrates the connection between our work and previous ideas put forward in the literature. A very interesting problem in quantum field theory is the search of quantities that vary monotonically along the RG flow, and thus can be used to quantify the number of degrees of freedom at any given scale. This has led to the formulation of $c-$theorem in 2D~\cite{Zamolodchikov:1986gt}, the $F-$theorem in 3D~\cite{Casini:2012ei}, and the $a-$theorem~\cite{Komargodski:2011vj} in 4D. At some point, it was conjectured that the thermal free energy could behave monotonically along a RG flow~\cite{Appelquist:1999hr,Appelquist:1999vs}. However, the model discussed in this section provides a counterexample to this conjecture, because it is known that the thermal free energy decreases along the RG trajectory that connects the UV fixed point corresponding to the critical $O(N)$ model to the IR fixed point describing $N-1$ free, massless bosons. 

\begin{figure}
    \centering
    \includegraphics[width=0.75\linewidth]{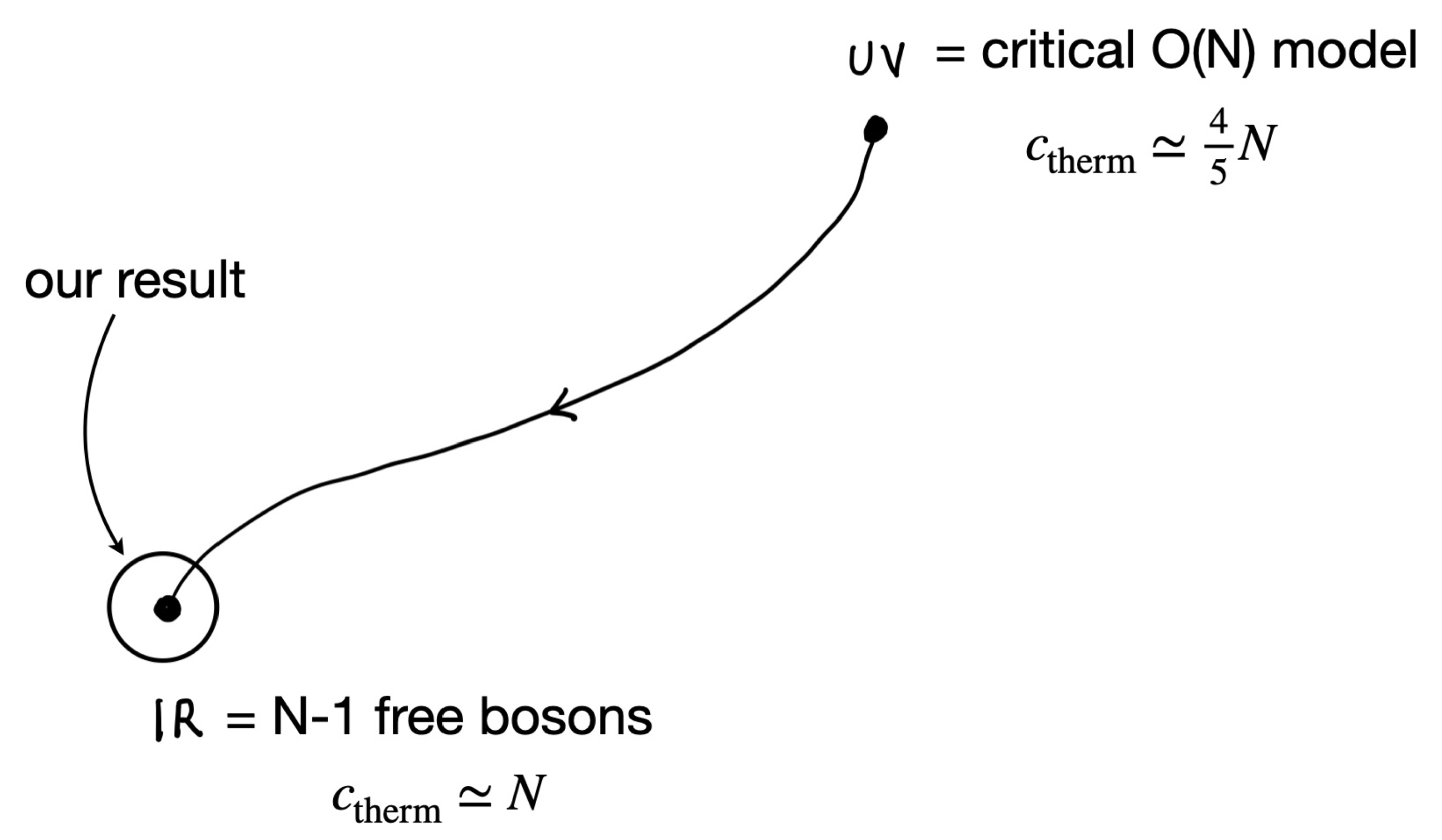}
    \caption{RG flow from the critical $O(N)$ model to a theory of free massless bosons.}
    \label{fig:RG flow}
\end{figure}

More precisely, parametrizing the thermal part of the free energy density at these fixed points as 
\begin{align}
	 \frac{\Omega^*}{V} = -\frac{1}{2}\frac{\zeta(3)}{\pi} c_{\rm therm} T^3 \ ,
\end{align}
the expectation was that $c_{\rm therm}$ would parametrize the effective number of degrees of freedom, and thus decrease along the RG flow. Instead, working in the large $N$ limit, we see from Eq. \eqref{eq: grand potential O(N) model} that $c_{\rm therm} \simeq N$ at the IR fixed point, while it was found that $c_{\rm therm} \simeq \frac{4}{5} N$ at the UV one~\cite{Chubukov:1993aau, Sachdev:1993pr}. 

Figure \ref{fig:RG flow} shows that the statement that irrelevant corrections must decrease the free energy density is much more limited in scope compared to the conjecture above, as it only concerns itself with what happens in a neighborhood of the IR fixed point. Even though, as we move away from the IR fixed point, the free energy density ultimately increases along the trajectory shown in Fig. \ref{fig:RG flow}, it still decreases sufficiently close to the IR fixed point.

\subsection{The \(T\bar{T}\) Deformation of the 2D Ising CFT}

The examples considered thus far concern irrelevant operators generated by integrating out heavy degrees of freedom. We now turn to an interesting and more subtle case: the $T\bar{T}$ deformation of the 2D Ising CFT. This example provides a novel test of our ideas. This is because the $T\bar{T}$ deformation is generated by coupling the original CFT to 2D gravity and integrating out the metric itself~\cite{Dubovsky:2017cnj} (which in 2D does not propagate any massless degrees of freedom), rather than by integrating out some heavy degrees of freedom. This distinction makes the $T\bar{T}$ deformation an intriguing laboratory for our conjecture. We will demonstrate that this deformation leads to a negative correction to the free energy only for one sign of the coupling constant. Notably, this is the same sign selected by a number of independent arguments in the literature based on causality, stability, and holography. 

\subsubsection{The Operator Spectrum and the Least Irrelevant Deformation}

The infrared (IR) fixed point of the 2D Ising model is described by the \(c=1/2\) unitary minimal model of Virasoro algebra \cite{DiFrancesco:1997nk}. This CFT has a sparse spectrum of primary operators. The low-dimension scalar primary operators are the identity \(\mathbb{I}\) (with scaling dimension \(\Delta=0\)), the spin operator $\sigma$ (with \(\Delta = 1/8\)) and the energy operator \(\epsilon\) (with \(\Delta=1\)). Both of these are relevant operators as their dimension is less than the spacetime dimension \(d=2\).

The theory contains no other scalar primary operators until much higher dimensions. Crucially, there are no scalar primaries with dimensions in the range \(2 < \Delta < 4\). The first available scalar operator that is irrelevant (\(\Delta > 2\)) is a composite operator built from the stress-energy tensor, \(T_{\mu\nu}\). This operator, commonly denoted \(T\bar{T}\), is defined as
\begin{eqnarray}
    T\bar{T}(x) \equiv \det(T_{\mu\nu}(x)).
\end{eqnarray}
It is a universal feature of any 2D quantum field theory that this operator has a scaling dimension of exactly \(\Delta=4\). Given the absence of other operators in the critical window, the \(T\bar{T}\) operator constitutes the least irrelevant scalar deformation for the 2D Ising CFT.

\subsubsection{Correction to the Thermal Free Energy}

We now consider deforming the Ising CFT action by the following operator:
\begin{eqnarray}
    S = S_{\text{Ising CFT}} + g \int d^2x \, T\bar{T}(x),
\end{eqnarray}
where \(g\) is a coupling constant with mass dimension \([g] = [M]^{-2}\). The first-order correction to the thermal free energy density is given by the thermal expectation value of the perturbing operator in the unperturbed CFT:
\begin{eqnarray}
    \frac{\delta \Omega^*}{V} = g \langle T\bar{T} \rangle_{T}.
\end{eqnarray}
A remarkable result by Zamolodchikov \cite{Zamolodchikov:2004ce} showed that this expectation value factorizes into the product of the thermal expectation values of the energy density, \(\varepsilon = \langle T_{00} \rangle_{T}\), and the pressure, \(p = \langle T_{11} \rangle_{T}\).\footnote{In the rest frame of the thermal bath, $\langle T_{01} \rangle_{T} =0$.} In any 2D CFT, these quantities are universal and related to the central charge \(c\):
\begin{eqnarray}
    \varepsilon = p = \frac{\pi c}{6} T^2.
\end{eqnarray}
For the 2D Ising CFT, the central charge is \(c=1/2\). Combining these results, the expectation value is
\begin{eqnarray}
    \langle T\bar{T} \rangle_{T} = \left( \frac{\pi c}{6} T^2 \right)^2 = \left( \frac{\pi (1/2)}{6} T^2 \right)^2 = \frac{\pi^2}{144} T^4.
\end{eqnarray}
The correction to the free energy density is therefore
\begin{equation} \label{eq:free_energy_correction}
    \frac{\delta \Omega^*}{V} = g \frac{\pi^2}{144} T^4.
\end{equation}

\subsubsection{Implications for the Sign of the Coupling}

Our conjecture posits that the leading correction from an irrelevant operator should lower the free energy, i.e., \(\delta \Omega^* < 0\). Applying this conjecture to the result in Eq. \eqref{eq:free_energy_correction}, we find a direct constraint on the sign of the coupling \(g\). Since both \(T^4\) and the numerical factor are positive, a negative free energy correction requires:
\begin{eqnarray}
    g < 0.
\end{eqnarray}
Notably, the sign \(g < 0\) derived from our conjecture is the same one selected by other physical considerations.  We summarize the main arguments here:
\begin{itemize}
    \item \textbf{Causality and S-Matrix:} The \(T\bar{T}\) deformation modifies a theory's S-matrix by a universal energy-dependent phase. It was argued in \cite{Dubovsky:2017cnj} that for a theory to be causal, this phase must correspond to a time delay, not a time advance. This requirement selects the sign corresponding to \(g < 0\).

    \item \textbf{Holography:} In the context of the AdS/CFT correspondence, the \(g < 0\) deformation is dual to moving the holographic boundary to a finite radial cutoff inside the bulk, resulting in a well-behaved geometry. The \(g > 0\) deformation, in contrast, corresponds to a cutoff at a timelike radius, which introduces pathologies such as closed timelike curves in the dual gravitational theory \cite{McGough:2016lol}.

    \item \textbf{Thermodynamics and Causality of Sound:} A thermodynamic argument based on the causality of sound propagation also singles out the same sign. As argued in \cite{Delacretaz:2021ufg}, in an effective field theory description of a CFT, one must consider the combined effect of all irrelevant operators on the speed of sound, \(c_s\). While irrelevant operators with dimension \(2 < \Delta < 3\) correctly push the speed of sound to be subluminal (\(c_s < 1\)), operators with dimension \(3 < \Delta < 4\) have the opposite effect and would naively lead to a superluminal, acausal sound speed. To prevent this, the \(T\bar{T}\) operator, which is generically present, must provide a compensating contribution. This physical requirement that causality be preserved at all scales fixes the sign of the \(T\bar{T}\) coupling to be \(g < 0\).
\end{itemize}
To summarize, the \(T\bar{T}\) deformation of the 2D Ising model provides a non-trivial case study. The calculation of the thermal free energy correction, when combined with our conjecture, uniquely selects the sign of the coupling \(g < 0\). This finding is in agreement with a strong consensus in the field, thereby lending additional support to the proposed conjecture.

\section{Cautionary Tales}
\label{sec:cautionary}
We now consider several cases in which an entropy positivity bound is not naively applicable to a low energy theory and we discuss why such a bound does not hold.  In particular, we emphasize that we only expect our bounds to hold when the ``reference" theory is an IR conformal fixed point and the ``target" theory contains irrelevant operators that break the conformal invariance of the reference theory.  In such cases, the entropy positivity bounds apply only to the leading irrelevant operators.  As we will show below, when these assumptions are violated, the naive entropy bounds do not hold.

\subsection{Conformal Superfluid}
As a first example, we consider the $(2+1)$D conformal superfluid.  Positivity bounds for EFT coefficients in this theory were derived in \cite{Creminelli:2022onn} using novel considerations for Lorentz violating theories and one might wonder if the entropy positivity bounds proposed here have any overlap with those previously derived. The effective action for a conformal superfluid in $(2+1)$D up to next-to-leading order in the derivative expansion is~\cite{Creminelli:2022onn}:
\begin{equation}
S= \int d^3 x \left[ \frac{c_1}{6}|\partial \chi|^3-2 c_2 \frac{(\partial|\partial \chi|)^2}{|\partial \chi|}+c_3\left(2 \frac{\left(\partial^\mu \chi \partial_\mu|\partial \chi|\right)^2}{|\partial \chi|^3}+\partial_\mu\left(\frac{\partial^\mu \chi \partial^\nu \chi}{|\partial \chi|^2}\right) \partial_\nu|\partial \chi|\right) \right] \ . \label{eq: action 3d conformal superfluid}
\end{equation}
where $|\partial\chi| \equiv \sqrt{-\partial_\mu\chi \partial^\mu\chi}$. One might expect entropy bounds to put constraints on the coefficients $c_2$ and $c_3$ of the leading irrelevant operators, as was done in \cite{Creminelli:2022onn}.

The spontaneous breaking of the Lorentz symmetry is generated once again by a time-dependent field configuration for the field $\chi$:
\begin{equation}
\chi(x)=\mu t+\pi(x)
\end{equation}
Expanding the action \eqref{eq: action 3d conformal superfluid} up to quadratic order in momentum space gives, in Fourier space,  
\begin{equation} \label{eq: quadratic pi Lagrangian}
{\cal L}^{(2)}=  \frac{c_1 \mu }{2} \ \tilde{\pi}(-\omega, -\boldsymbol{k})\left[\omega^2-c_s^2 \boldsymbol{k}^2+\frac{4\left(c_2+c_3\right)}{c_1 \mu^2} \omega^2\left(\omega^2-\boldsymbol{k}^2\right)\right] \tilde{\pi}(\omega, \boldsymbol{k})
\end{equation}
where $c_s=\frac{1}{\sqrt{2}}$, as appropriate for a conformal superfluid in (2+1)D. From this, we can easily calculate the dispersion relation perturbatively in $k/\mu$ to find 
\begin{equation}
    \omega (k) \simeq c_s k \left[ 1-\frac{2\left(c_2+c_3\right)}{c_1 \mu^2} \left(c_s^2-1\right)k^2 \right]
\end{equation}

To invoke our entropy bound, we must consider the leading corrections to the free energy density due to the presence of the leading irrelevant operators.  Naively, this would appear to come from the $c_2 + c_3$ term in the quadratic Lagrangian~\eqref{eq: quadratic pi Lagrangian}.  To find the corresponding change in the free energy density, we use the (2+1)D analog of the expression \eqref{eq: thermal free energy free (3+1)D} for the thermal free energy of free bosons with dispersion relation $\omega (k)$.  It is straightforward to expand in powers of $k/\mu$ and obtain the leading correction to $\Omega^*$ at small temperatures, $T \ll \mu$: 
\begin{align}
	\frac{\Omega^*}{V} \simeq - \frac{\zeta(3)}{\pi \beta^3} +\frac{c_2+c_3}{c_1 \mu^2 } \frac{48\, \zeta(5)}{\pi \beta^5} \ .
\end{align}
Self-interactions would have contributed corrections of $\mathcal{O}(T^4/\mu^4)$. Thus, we find that corrections to the thermal part of the free energy density are negative provided $c_2 + c_3 <0$ since $c_1$ must be positive to avoid a ghost instability. However, coming to the conclusion that this inequality must be satisfied would be too quick; the deformations parametrized by $c_2$ and $c_3$ do not move us away from a CFT, but instead, it moves us around the space of CFTs. This action is explicitly built with the goal of realizing conformal symmetry. Had these operators not belonged to these class of theories, but instead deformed away from conformality, then our bounds would have been applicable. In fact, \cite{Creminelli:2022onn} considers a bona-fide UV completion where $c_2,c_3>0$, showcasing the breakdown of positivity bounds for arbitrary operators.

\subsection{The $\lambda \phi^4$ Theory}

At first glance, a massless scalar field theory with a quartic self-interaction in four dimensions, described by the Lagrangian
\begin{align}
	\mathcal{L} = -\frac{1}{2}(\partial_\mu \phi)^2 - \lambda\phi^4, \label{eq:phi4_lagrangian}
\end{align}
appears to be a prime candidate for testing our conjecture. The free theory ($\lambda=0$) is a conformal field theory, and the $\phi^4$ operator, with scaling dimension $\Delta=4$, is a marginally irrelevant deformation. One might therefore expect that the sign of the coupling $\lambda$ would be constrained by the requirement that the thermal free energy must decrease, $\Delta\Omega^* < 0$. Let's explore the consequences of this line of reasoning. We'll see that our conjecture must apply to purely irrelevant rather than marginally irrelevant deformations. Along the way, we will also develop a clearer understanding of the difference between our requirement and the typical stability requirements.

The leading-order correction to the thermal grand potential density due to the interaction term can be calculated using standard thermal field theory methods. In the high-temperature limit, this correction is dominated by the thermal fluctuations of the field. The change in the grand potential density is given by the expectation value of the interaction part of the Lagrangian, evaluated in the free theory:
\begin{align}
	\frac{\Delta\Omega^*}{V} = \lambda \langle \phi^4(x) \rangle_T^*.
\end{align}
The thermal expectation value of $\phi^4$ can be computed using Wick's theorem, which gives $\langle \phi^4 \rangle_T = 3 (\langle \phi^2 \rangle_T)^2$, where $\langle \phi^2 \rangle_T$ is the thermal variance of the field at a point. The temperature-dependent part of this propagator for a massless scalar field is a classic result (see, e.g., \cite{Kapusta:2006pm}):
\begin{align}
	\langle \phi^2(x) \rangle_T^* = \int \frac{d^3k}{(2\pi)^3} \frac{1}{k} \frac{1}{e^{\beta k} - 1} = \frac{T^2}{12}.
\end{align}
Combining these results, the leading correction to the thermal grand potential density is
\begin{align}
	\frac{\Delta\Omega^*}{V}  = \frac{\lambda T^4}{48}. \label{eq:phi4_omega_corr}
\end{align}
According to our conjecture, we expect $\Delta\Omega^* < 0$, which, based on Eq. \eqref{eq:phi4_omega_corr}, directly implies that the coupling must be negative: $\lambda < 0.$

This might appear to be in tension with the fact that a theory with $\lambda < 0$ is unstable, with a potential that is unbounded from below. However, this apparent conflict highlights a crucial distinction regarding the physical origin of the effective operators we are bounding. While the derivative operators considered in previous sections have their signs constrained by causality and the analyticity of the S-matrix, the $\phi^4$ operator is constrained by vacuum stability. Crucially, the sign required for stability ($\lambda > 0$) is not the sign that is naturally generated when integrating out a heavy particle at tree level.

To see this explicitly, consider a simple UV completion where our light scalar $\phi$ interacts with a heavy scalar field $\Phi$ of mass $M$ via a cubic coupling:
\begin{align} \label{eq: UV Lagrangian lambda phi^4}
	\mathcal{L}_{UV} = -\frac{1}{2}(\partial_\mu \phi)^2 - \frac{1}{2}(\partial_\mu \Phi)^2 - \frac{1}{2}M^2\Phi^2 - \frac{g}{2}\phi^2 \Phi.
\end{align}
In the low-energy effective theory ($E \ll M$), the heavy field $\Phi$ can be integrated out. At tree-level, this is equivalent to solving for $\Phi$ using its classical equation of motion, $(\partial^2 + M^2)\Phi = - (g/2) \phi^2$, in the low-energy limit where the derivative term can be neglected: $\Phi(x) \approx -\frac{g}{2M^2}\phi^2(x)$. Substituting this solution back into the potential part of the Lagrangian \eqref{eq: UV Lagrangian lambda phi^4} yields the tree-level effective potential for $\phi$:
\begin{align}
	V_{\rm eff}(\phi) &\approx \frac{1}{2}M^2 \left(-\frac{g}{2M^2}\phi^2\right)^2 + \frac{g}{2}\phi^2 \left(-\frac{g}{2M^2}\phi^2\right) =  -\frac{g^2}{8M^2}\phi^4.
\end{align}
Comparing this to the potential in our model, $V(\phi) = \lambda\phi^4$, we identify the Wilson coefficient as $\lambda = -g^2/(8M^2)$. Since $g^2$ and $M^2$ are positive, the coefficient $\lambda$ generated via tree-level scalar exchange is necessarily negative. This result is quite general; the exchange of a single boson at tree level mediates an attractive force, corresponding to a negative potential energy contribution.

This example provides a sharp illustration of how our proposed entropy criterion is distinct from the established concepts of both vacuum and thermodynamic stability. The three conditions impose mutually exclusive constraints on the coupling $\lambda$:
\begin{enumerate}
	\item \textbf{Vacuum Stability} requires $\lambda > 0$ to ensure the potential is bounded from below at $T=0$.
	\item \textbf{Thermodynamic Stability} requires the heat capacity $c_V = T \partial s / \partial T$ to be positive, which imposes a perturbative upper bound, $\lambda < 8\pi^2/15$.
	\item \textbf{Our Conjecture} requires $\Delta\Omega^* < 0$, which implies $\lambda < 0$.
\end{enumerate}
The fact that our conjecture aligns with the sign produced by the simplest tree-level UV completions ($\lambda < 0$), while diverging from the requirements of both vacuum and thermodynamic stability, provides strong evidence that it probes a different aspect of physical consistency—one tied to the causal generation of operators rather than the stability of the resulting state.

\section{Summary and Discussion}
\label{sec:discussion}

In this work, we put forward a conjecture for constraining low-energy effective field theories that is rooted in thermodynamic consistency. In particular, we posited that irrelevant deformations of a low energy CFT whose Wilson coefficients can be generated by integrating out a weakly-coupled UV theory must give a negative contribution to the thermal part of the grand potential: $\Delta\Omega^*(\beta,\mu) < 0$.  We derived that, under general thermodynamic principles, a negative shift in the grand potential at fixed temperature and chemical potential is equivalent to a positive shift in the entropy at fixed energy and charge, $\Delta S(E,Q) > 0$. This provides a consistency criterion that, unlike many positivity bounds derived from S-matrix analyticity, does not rely on Lorentz invariance.

We demonstrated that conjecture was supported by a number of well known effective field theories. We found that our entropy bound correctly reproduced the known positivity constraints for the $(\partial\phi)^4$ Goldstone EFT, both in a Lorentz invariant setup and at finite chemical potential, as well as for the Euler-Heisenberg action. Furthermore, we showed that it is consistent with the known sign of the leading thermal correction in the $O(N)$ nonlinear sigma model in $(2+1)D$ and for $N > 2$ where the sign of the correction is dictated by symmetry arguments, showing self-consistency of the theory with respect to our bounds.  Moreover, our bounds selected the physically correct sign of the coupling for the $T\bar{T}$ deformation of a 2D CFT. This last example is especially significant, as the deformation arises not from integrating out heavy particles, but from coupling the theory to 2D gravity and integrating out the non-propagating metric. This suggests that our conjecture may apply to a broader class of tree-level completions than initially assumed and probing the boundaries of this conjecture would be of great interest. It would be important to provide a non-perturbative formulation of our core assumptions and to investigate the existence of a proof that could also apply to strongly coupled UV completions.

This work opens other promising avenues for future investigation. First, there are a number of additional test cases we wish to examine. For instance, we plan next to apply our constraints to massless chiral perturbation theory, to a theory of chiral fermions with a four-fermi interaction, or to a $P(X)$ superfluid EFT.  These examples would provide further stringent tests of our conjecture. However, these models present technical challenges related to the careful handling of IR divergences, which go beyond the scope of the present analysis.

Second, amplitude-based techniques can provide bounds beyond the leading-order operators. The standard positivity bounds on the Wilson coefficient of the $s^2$ term in a forward scattering amplitude can be extended by considering higher-order derivatives with respect to the Mandelstam variable $s$ at $t=0$. This way, one can derive an infinite tower of sum rules, often known as ``extended positivity bounds,'' that constrain linear combinations of the coefficients of higher-dimension operators~\cite{deRham:2017avq,deRham:2017zjm}. It is not immediately obvious if or how our entropy bound could be extended to yield such a tower of constraints.  It would be interesting to explore whether some property of the thermal partition function, perhaps related to its analytic structure in the complex temperature plane~\cite{Fisher:1971ks}, could play a role analogous to the analyticity of the $S$-matrix, providing a systematic way to bound the entire spectrum of irrelevant operators.

\acknowledgments
We would like to thank Ira Rothstein for useful discussions. RP would also like to thank the participants of the workshop ``Constraining Effective Field Theories without Lorentz'' (IFPU, July 7-11, 2025), and in particular Paolo Creminelli, Diptimoy Ghosh, Lam Hui, Joan Elias Mir\'o, and Alessandro Podo for feedback and stimulating discussions. This work is supported by the US Department of Energy grant DE-SC0010118.  LF is partially supported by the Rafael del Pino Foundation.

\bibliographystyle{JHEP}
\addcontentsline{toc}{section}{References}
\bibliography{entropybounds}

\end{document}